\documentclass[12pt,preprint]{aastex}
\usepackage{epsf}

\shorttitle{The M~87 Jet During the VHE Flare in April 2010}
\shortauthors{Hada et al.}

\begin{document}

\title{VLBI Observations of the Jet in M~87 During the Very-High-Energy
$\gamma$-ray Flare in April 2010}

\author{Kazuhiro Hada\altaffilmark{1,2,3}, Motoki Kino\altaffilmark{3}, Hiroshi
Nagai\altaffilmark{3}, Akihiro Doi\altaffilmark{4}, Yoshiaki
Hagiwara\altaffilmark{2,3}, Mareki Honma\altaffilmark{2,3}, Marcello
Giroletti\altaffilmark{1}, Gabriele Giovannini\altaffilmark{1,5} and Noriyuki
Kawaguchi\altaffilmark{2,3}}

\affil{$^1$INAF Istituto di Radioastronomia, via Gobetti 101, I-40129 Bologna,
Italy}

\affil{$^2$Department of Astronomical Science, The Graduate University for
Advanced Studies (SOKENDAI), 2-21-1 Osawa, Mitaka, Tokyo 181-8588, Japan}

\affil{$^3$National Astronomical Observatory of Japan, Osawa, Mitaka, Tokyo
181-8588, Japan}

\affil{$^4$Institute of Space and Astronautical Science, Japan Aerospace
Exploration Agency, 3-1-1 Yoshinodai, Chuo, Sagamihara 252-5210, Japan}

\affil{$^5$Dipartimento di Astronomia, Universit\`a di Bologna, via
Ranzani 1, I-40127 Bologna, Italy}

\begin{abstract}
 We report on the detailed radio status of the M~87 jet during the
 Very-High-Energy (VHE) $\gamma$-ray flaring event in April 2010,
 obtained from high-resolution, multi-frequency, phase-referencing VLBA
 observations. We especially focus on the properties for the jet base
 (the radio core) and the peculiar knot HST-1, which are currently
 favored as the $\gamma$-ray emitting sites. During the VHE flaring
 event, the HST-1 region remains stable in terms of its structure and
 flux density in the optically thin regime above 2~GHz, being consistent
 with no signs of enhanced activities reported at X-ray for this
 feature. The radio core shows an inverted spectrum at least up to
 43~GHz during this event. Astrometry of the core position, which is
 specified as $\sim$$20~R_{\rm s}$ from the central engine in our
 previous study, shows that the core position is stable on a level of
 4~$R_{\rm s}$.  The core at 43 and 22~GHz tends to show slightly
 ($\sim$10\%) higher flux level near the date of the VHE flux peak
 compared with the epochs before/after the event. The size of the 43-GHz
 core is estimated to be $\sim$17~$R_{\rm s}$, which is close to the
 size of the emitting region suggested from the observed time scale of
 rapid variability at VHE. These results tend to favor the scenario that
 the VHE $\gamma$-ray flare in 2010 April is associated with the radio
 core.
\end{abstract}


\section{Introduction}
The location and the physical properties of Very-High-Energy (VHE) $\gamma$-ray
emission from relativistic jets are one of the most intriguing questions in
astrophysics. The nearby radio galaxy M 87 is a well-known VHE $\gamma$-ray
emitter since the first identification of TeV emission by HEGRA in 1998/1999
\citep{aharonian2003}. Thanks to its proximity \citep[$D=16.7$~Mpc,
$z=0.00436$;][]{jordan2005} and a large black hole mass~\citep[$M_{\rm BH} \simeq
(3-6) \times 10^9~M_{\odot}$;][]{macchetto1997, gebhardt2009}\footnote{In this
paper we adopt $M_{\rm BH} = 6\times 10^9M_{\odot}$ along with \citet{hada2011},
although we note that the exact value of $M_{\rm BH}$ in M~87 is still
controversial and should be carefully considered. One can re-scale the values in
$R_{\rm s}$ unit in this paper by multiplying a factor of 2 if $M_{\rm BH}=3\times
10^9M_{\odot}$ is used.}, the jet structure can be resolved under 100
Schwarzschild radii scale ($R_{\rm s}$) with VLBI observations~\citep[][hereafter
H11]{junor1999, ly2007, asada2012, hada2011}, providing a unique opportunity to
probe the connection of the VHE $\gamma$-ray with the relativistic jet by
isolating the detailed substructures.

Recently there have been three remarkable VHE flares from M~87; the events in
2005, 2008 and 2010. In the 2005 event~\citep{aharonian2006}, the VHE flare was
accompanied by radio-to-X-ray flares from HST-1, a peculiar knot located at a
de-projected distance of at least $\sim$120~pc downstream of the
nucleus~\citep{harris2006}\footnote{The exact distance of HST-1 depends on the
viewing angle of the M~87 jet.  The observed superluminal motions of HST-1 suggest
a smaller viewing angle close to $\sim$15$^{\circ}$~\citep{biretta1999}. The
recent optical polarization study of HST-1 supports this
trend~\citep{perlman2011}.}, with the emergence of superluminal ($\sim$$4c$) radio
features~\citep{cheung2007}. These lead to the strong argument that the VHE
emission originates in HST-1~\citep[e.g.,][]{stawarz2006, cheung2007, harris2008,
harris2009}, although there are still some debates on this
interpretation~\citep[e.g.,][]{georganopoulos2005}. In the case of the 2008 event,
on the other hand, the \textit{Chandra X-ray Observatory} detected an enhanced
X-ray flux from the nucleus, while HST-1 maintained a comparatively constant
flux. In addition, synchronized VLBA observations at 43~GHz revealed a strong flux
increase from the radio core that lasted the subsequent $\sim$two months. These
provide evidence that the VHE flare in 2008 originates in the
core~\citep{acciari2009}.

The third one occurred more recently, in April 2010, where the VHE flare has been
clearly detected during the joint monitoring campaign by H.E.S.S, VERITAS and
MAGIC~\citep[][hereafter A12]{ong2010, aliu2012, abramowski2012}. The detected
flare displays a smooth rise and decay in flux with a peak around MJD 55296 (2010
April 9-10), reaching a historic high state of about 20\% of the flux of the Crab
Nebula. Interestingly, \textit{Chandra} observations taken $\sim$3 days after the
peak of the VHE flare detected an enhanced flux from the nucleus, whereas HST-1
remained in a low state~\citep{harris2011}. Short time scales of variabilities
observed at VHE and X-ray suggest the size of the emitting region to be of the
order of a few light days times the Doppler factor $\delta$, which corresponds to
less than $\sim$10$\delta~R_{\rm s}$.  We note, however, that a detailed study on
HST-1 by \citet[][hereafter G12]{giroletti2012} confirmed a recurrence of this
structure and its possible connection with the VHE activity.
  
While the broad band light curve is studied in A12, in this letter we report on
the detailed radio status of the M~87 jet during the 2010 VHE flaring event
obtained from high-resolution VLBA observations. We especially focus on the
multi-frequency observations which succcessfully synchronized with this event,
providing a wealth of information for HST-1 and the inner jet during the flare on
milliarcsecond (mas) scale. The data and the analysis are described in the next
section. We then show the results in section~3. In the final section, we discuss
the obtained results and make a summary. In the present paper, the spectral index
$\alpha$ is defined as $S(\nu)\propto \nu^{+\alpha}$.

\section{Observations and Data Reduction}
M~87 was observed with the VLBA at 2, 5, 8, 15, 22 and 43~GHz on April 8 and 18
2010 (MJD~55294 and 55304), which are just before and after the date of the
maximum flux of the VHE flare (MJD~55296). These data, which were not presented in
A12 except for the results of the HST-1 flux at 2~GHz, are identical to those
presented in~H11, where we investigated the core shift of M~87 using the
phase-referencing technique relative to the nearby radio source M~84. The details
of the observations and the data reduction processes including the astrometric
analysis are described in H11. For this paper we partly re-analysed the data in
order to properly examine the radio status of M~87 including the HST-1 region,
which is located $\sim$900~mas away from the phase-tracking center.  The data were
averaged only to 5~s in time and kept individual channels (1~MHz width) separated
before imaging process to avoid time/bandwidth smearing effects at the location of
HST-1.

In addition, we also analysed many available VLBA archival data at 43 and 22~GHz
to investigate the lightcurve of the inner jet region around the VHE flare (see
Section 3.2). These consist of the data on January 18 (22~GHz), April 4 (22~GHz),
May 1 (22~GHz), May 15 (22~GHz) and May 30 (22 and 43~GHz) 2010, which were not
included yet in A12.

\begin{figure*}[htbp]
    \centering \includegraphics[angle=0,width=1.0\textwidth]{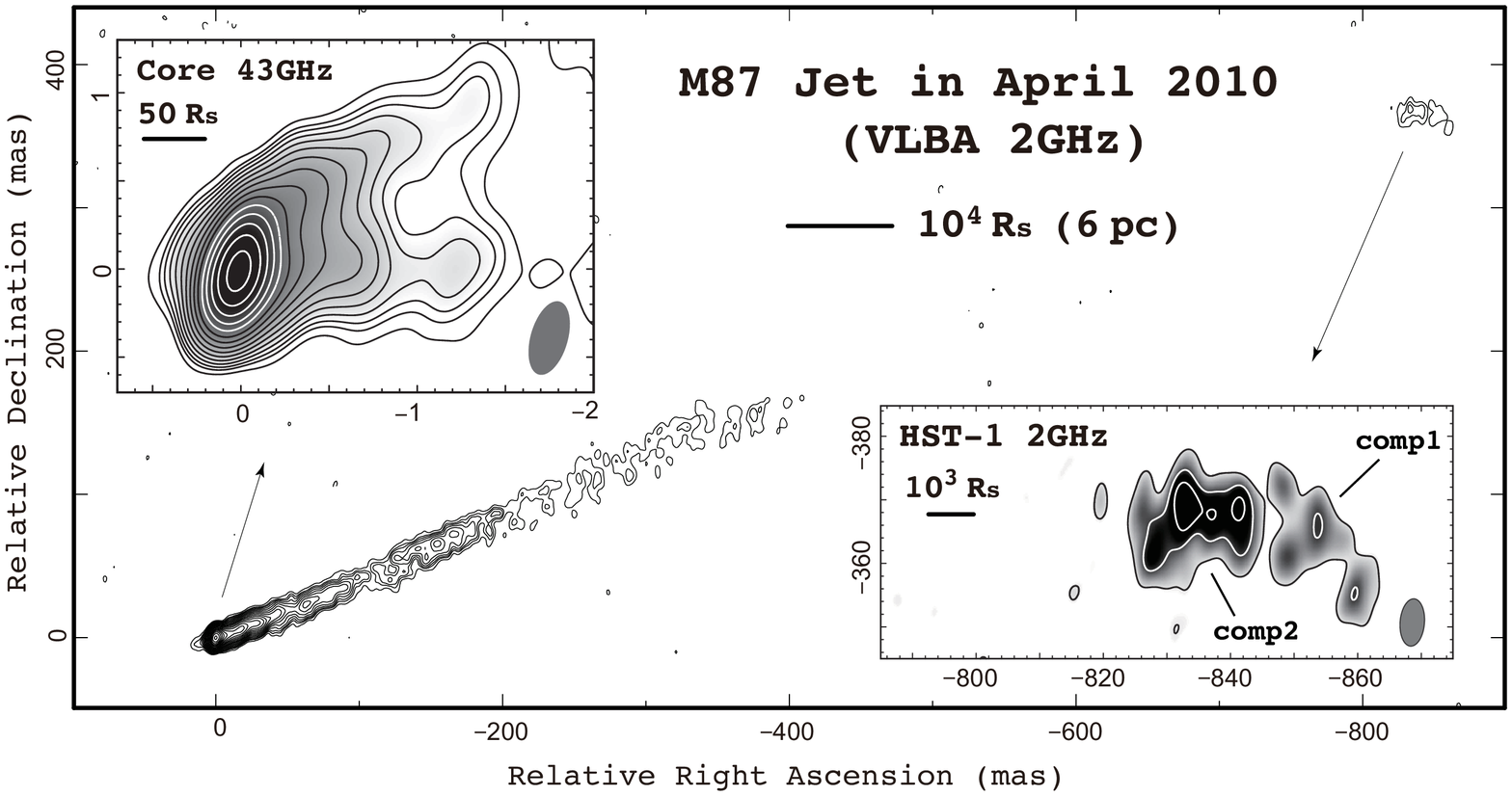}
    \caption{A summary of VLBA images of the M~87 jet during the VHE
    $\gamma$-ray flare in April 2010. The main (global) image at 2~GHz
    was obtained by combining the data on April 8 and 18 2010. The
    bottom right inset indicates a close up view toward the HST-1
    region. The nomenclatures of two main features (comp1 and comp2) are
    based on G11. The upper left inset indicates a 43-GHz image for the
    core and the inner jet (also obtained by averaging both epochs). The
    beam sizes at 2/43~GHz are 7.5$\times$3.9 mas in P.A. $-5^{\circ}$
    (bottom right in the inset of HST-1) and 0.43$\times$0.21 mas in
    P.A. $-16^{\circ}$ (bottom right in the 43-GHz image),
    respectively. For each image, contours start from $-$1, 1,
    2... times 3$\sigma$ image rms levels ($3\sigma =
    1.0/3.3$~mJy~beam$^{-1}$ at 2/43~GHz) and increasing by factors of
    1.4.} \label{fig:image}
   \end{figure*}

Images were created in DIFMAP software with iterative phase/amplitude
self-calibration. Several weighting schemes were used depending on the target
region.  In Figure~\ref{fig:image} we summarize representative images of the M~87
jet obtained by our observations.

\section{Results}
\subsection{The HST-1 Region}
The HST-1 region was detected on both epochs at 2 and 5~GHz on a level of
12$\sigma$ and 7$\sigma$ image rms, respectvely. At 8~GHz, the analysis combined
for both epochs with a relatively strong $uv$-tapering detected this region with
$\sim$8$\sigma$. The feature was not detected at 15, 22 and 43~GHz due to the
image sensitivity limit. A part of the HST-1 properties in these epochs (i.e. the
distance from the core and the position angle of the HST-1 region at 2~GHz) is
already reported in G12 in the context of the long-term kinematic study of this
feature.

The overall structure observed in these epochs is similar to that on 2010.07, the
image for which is presented in G12. The HST-1 region is resolved into two main
subfeatures with an overall extension of $\sim$40~mas at 2~GHz. Model fitting with
two Gaussian components yields the sizes of these features as $\sim$18~mas
(1.5~pc, 2.5$\times10^3$~$R_{\rm s}$) and $\sim$16~mas (1.3~pc,
2.2$\times10^3$~$R_{\rm s}$) for the upstream/downstream components, respectively.
While the emergence of a new component from HST-1 upstream is discovered in the
later epochs of 2010~(G12), such a feature is not found yet in our observations on
April 8 and 18. The brightness temperature for the brightest component is
estimated to be $\sim$$1\times 10^7$~K, which is similar to the upper limit of
$9\times10^6$~K derived in the previous study at 15~GHz~\citep{chang2010}
                   
\begin{figure}[htbp]
   \centering
    \includegraphics[angle=0,width=1.0\columnwidth]{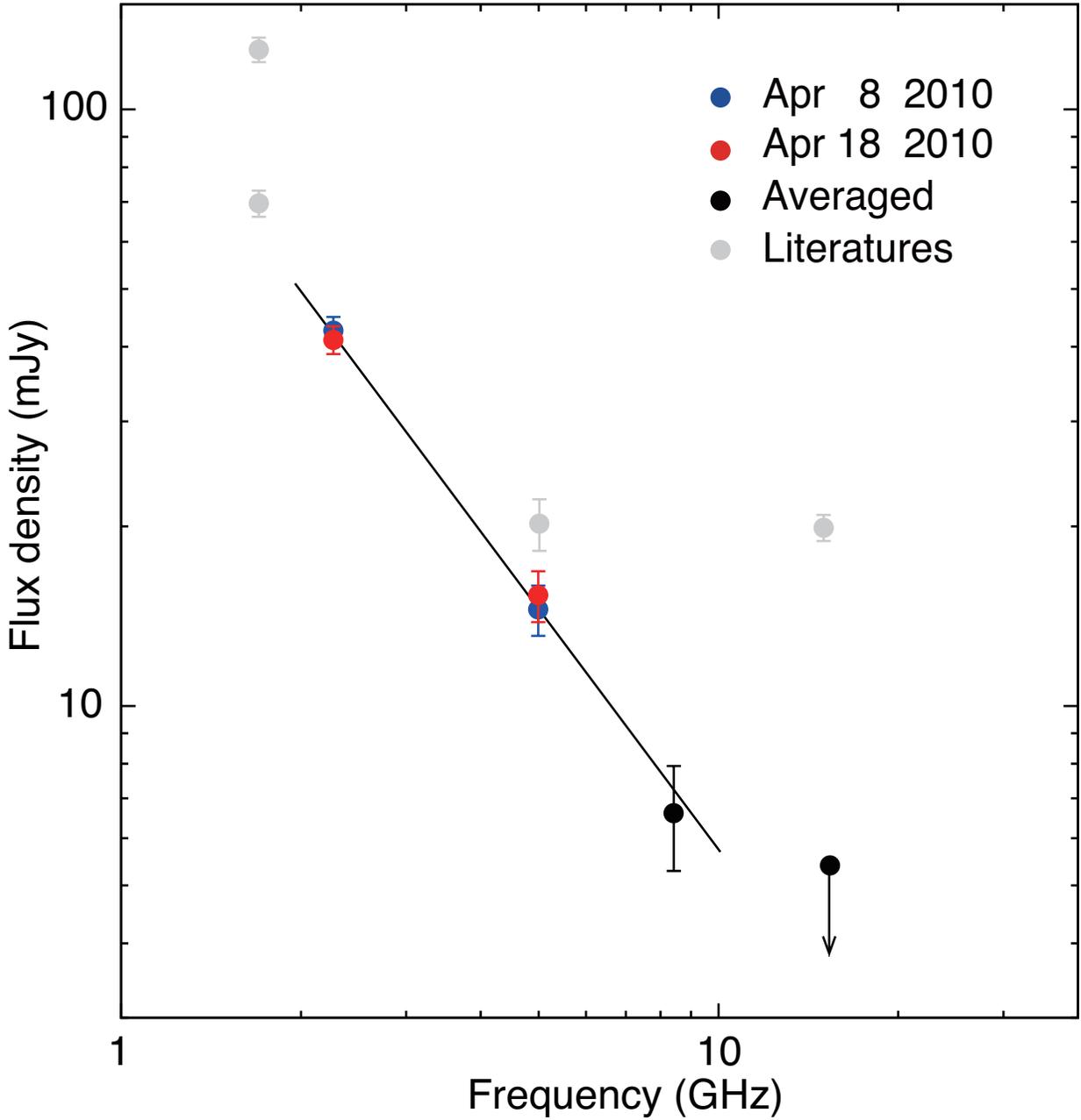}
    \caption{Integrated spectra for the HST-1 region. The blue/red
    circles at 2 and 5~GHz indicate the data on April 8/18 2010,
    respectively. The black circles indicate the flux density at 8~GHz
    and the upper limit at 15~GHz, which are obtained from the averaged
    image for the two epochs. A solid line indicates the averaged
    spectral shape for the two epochs. For reference, VLBI integrated
    flux densities in earlier epochs (grey circles) are provided from
    the literatures: 2005.82 and 2009.64 at 1.7~GHz, 2010.24 at 5~GHz,
    and 2005.85 at 15~GHz.}  \label{fig:hst1spec}
   \end{figure}

In Figure~\ref{fig:hst1spec}, we show the integrated radio spectrum of
the HST-1 region in these epochs. The HST-1 region shows a steep
spectrum with an averaged spectral index of $\alpha \sim -1.2$,
indicating that the emission region is optically thin. This radio
spectral index seems to be slightly steeper than the values for the
optical bands \citep[$\alpha_{\rm O-UV}\lesssim-0.7$;][]{perlman2011},
which were measured between 2002 and 2007. The spectral shapes are quite
similar between these two epochs and no significant flux variation was
found within the errors. The magnitudes of the flux densities in these
epochs appear to follow the long-term, monotonically-decaying trend that
continues since the maximum phase in 2005. These observational
characteristics at the radio frequencies are consistent with no signs of
enhanced activities in HST-1 at X-ray during
2010~\citep[A12;][]{harris2011}.

\subsection{The Core and the Inner Jet Region}
The inner jet region was clearly detected at all frequecies during the VHE flaring
phase. This region is characterized by the compact core with the edge-brightended
structure as seen in the 43-GHz image in Figure~\ref{fig:image}. Similar images
are obtained when images are created separately for April 8 and 18.

\begin{figure}[htbp]
    \centering \includegraphics[angle=0,width=1.0\columnwidth]{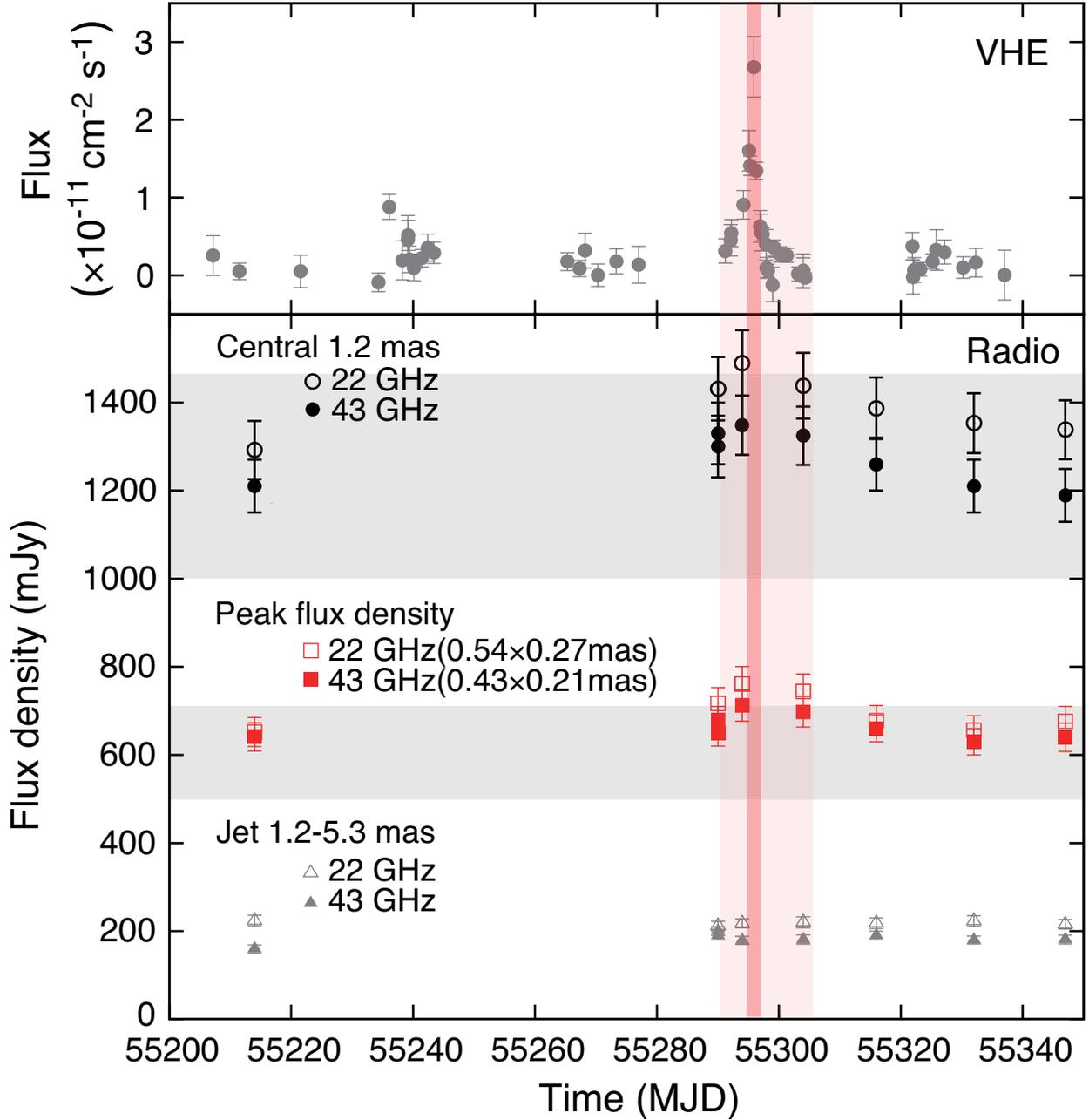}
    \caption{43- and 22-GHz lightcurves of the central region of M~87
    near the VHE flare in April 2010. The fluxes for three different
    regions (i), (ii) and (iii) are provided (see text for the detailed
    descriptions). The light red area indicates the period where the VHE
    $\gamma$-ray flaring event was covered by the VHE telescopes (see
    the top panel, which is taken from A12). The dark red area indicates
    the date of the VHE $\gamma$-ray flux peak (MJD~55296). For
    reference, we show the ranges of the flux densities at 43~GHz before
    the 2008 flare for the regions (i) and (ii), which represent typical
    flux ranges for the quiescent phase~\citep[][]{acciari2009}}
    \label{fig:core_curve}
   \end{figure}

In Figure~\ref{fig:core_curve}, we show the light curve of the inner jet region at
43 and 22~GHz around the 2010 flaring event. This is an updated version of the
43-GHz light curve presented in A12, where the data points analysed here were not
included yet. Based on \citet{acciari2009} and A12, the fluxes for three different
regions are provided; (i) the peak flux when convolved with a beam of $0.43 \times
0.21$~mas in P.A.$ -16^{\circ}$ at 43~GHz or $0.54 \times 0.27$~mas in P.A.$
-10^{\circ}$ at 22~GHz, (ii) the nucleus (the deconvolved flux in the circular
region with radius 1.2~mas = 0.1~pc, centered on the intensity peak), (iii) the
flux integrated along the jet between distances of 1.2 and 5.3~mas from the
intensity peak. We assign 5\% errors for all of the data points because the
amplitude calibration of the VLBA is typically accurate within this uncertainty.
As seen in Figure~\ref{fig:core_curve}, the inner jet region shows a relatively
moderate flux evolution around the VHE flare. Near the date of the VHE flux peak,
the innermost region (i) (and also (ii)) tends to be at a slightly higher flux
level ($\sim$10\% both at 22 and 43~GHz) compared with the other epochs. This is
in contrast to the event in 2008, where the 43-GHz core underwent a remarkable
flux increase (up to $\sim$30\%) lasting the subsequent $\sim$two months together
with a flux enhancement from the X-ray core~\citep{acciari2009}.

\begin{figure}[htbp]
     \centering
     \includegraphics[angle=0,width=1.0\columnwidth]{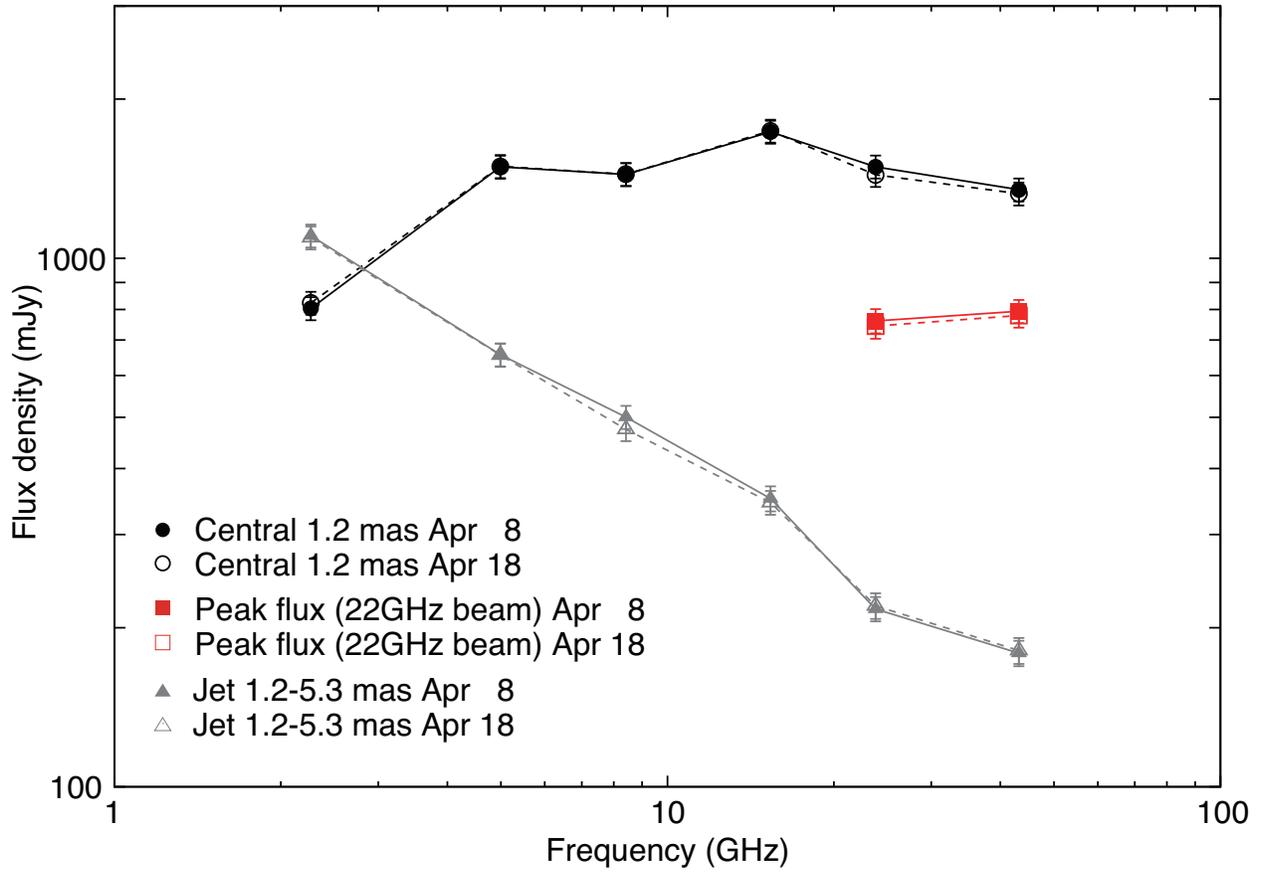}
     \caption{Radio spectra of the central region of M~87 on April 8 and
     18 2010. The peak fluxes at 22 and 43~GHz were estimated with the
     synthesized beam at 22~GHz ($0.54\times0.27$~mas in
     P.A. $-10^{\circ}$).}  \label{fig:core_spec}
    \end{figure}

In Figure~\ref{fig:core_spec}, we next show the radio spectra of the inner jet on
April 8 and 18 for the three regions. Note that, regarding the peak fluxes at 22
and 43~GHz, the data have all been convolved with a 22-GHz beam to match the
spatial resolutions. The spectra of each region looks the quite similar between
April 8 and 18. We found the following spectral characteristics as a function of
the measured region; the spectra gradually change from steep to flat toward the
upstream side ($\alpha\sim-0.7$ at (iii) to $\sim-0.1$ at (ii)), and the spectrum
of the innermost region (i) becomes slightly inverted between 22 and 43~GHz with
$\alpha \sim 0.1$, indicating that region (i) is optically thick at these
frequencies. This is consistent with the detection of the core shift reported in
H11, because this effect occurs only when the core represents an opaque part of
synchrotron emission at each frequency. For some of other radio sources, their
inverted radio cores are caused by foreground free-free
absorption~\citep[e.g.,][]{kellermann1966, odea1998}, but that does not seem to be
the case for M~87~\citep{ly2007}. The measured frequency dependence of the core
shift $\nu^{-0.94\pm0.09}$ by H11 is in good agreement with $\nu^{-1}$, which is
typically expected when the core is dominated by
synchrotron-self-absorption~\citep[e.g.,][]{bk1979, konigl1981, lobanov1998}.

\begin{figure}[htbp]
     \centering
     \includegraphics[angle=0,width=1.0\columnwidth]{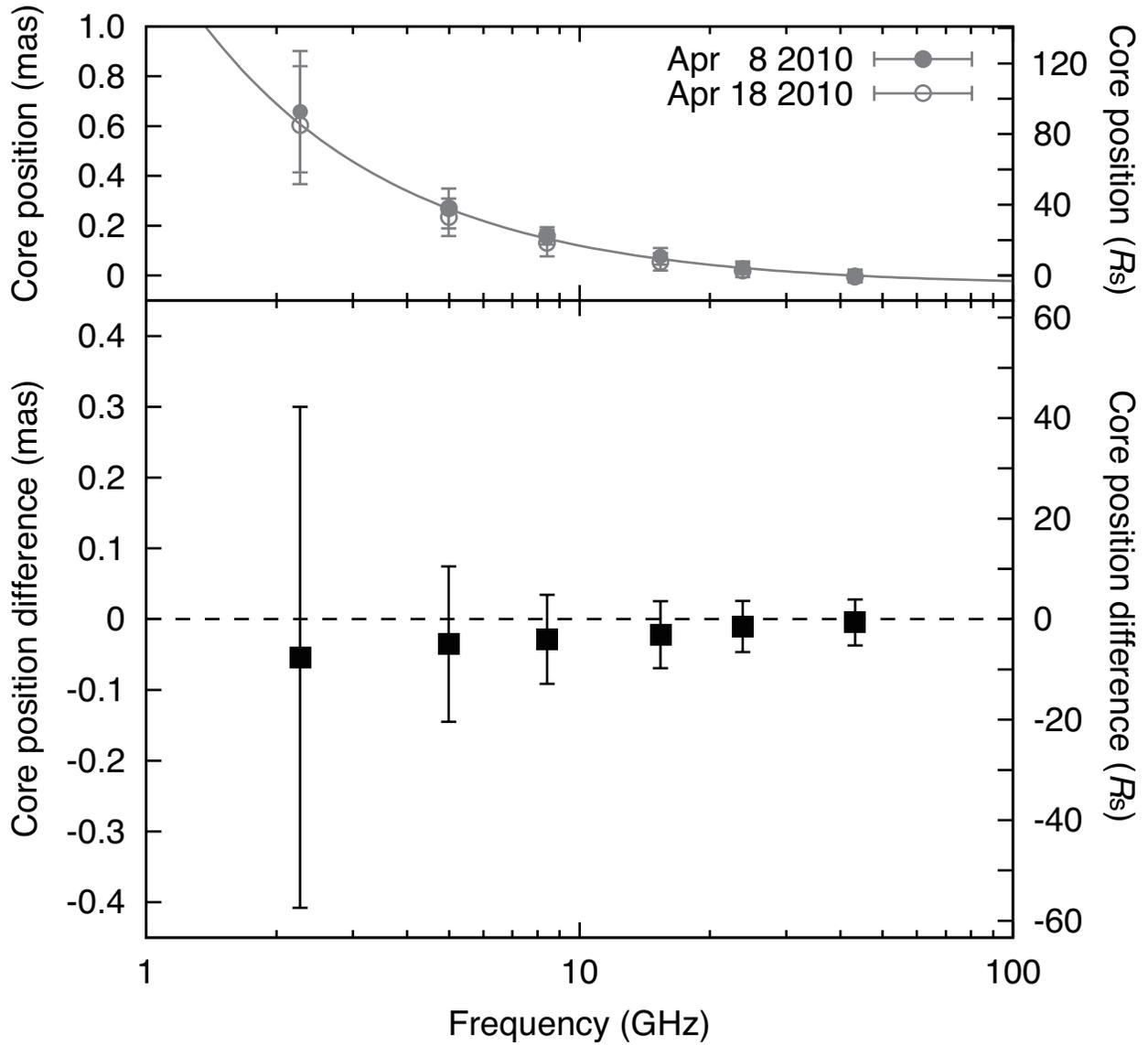}
     \caption{Astrometry of the core positions. (top panel) Detected
     core shifts in RA direction on April 8 and 18
     \citep{hada2011}. (bottom panel) Difference of the core position
     (in RA direction) between April 8 and April 18 at each frequency
     (the positions on April 18 minus the ones on April 8). Positive
     direction indicates the jet direction.}  \label{fig:core_pos}
    \end{figure}

In Figure~\ref{fig:core_pos}, we finally show the difference of the core shift (in
RA direction) between these two epochs (i.e., $r_{\rm core, \nu}(t={\rm
MJD~55304}) - r_{\rm core, \nu}(t={\rm MJD~55294})$).  Because the detected core
shifts on April 8 and 18 are quite similar each other, we did not find any
significant time evolution of the core position over the all observed
frequencies. The result at 43~GHz, where the measurement attains the highest
position accuracy, indicates that the core position remains stable on a level of
4~$R_{\rm s}$ (projected scale on the sky) during this period. A similar result is
reported during the VHE flare in 2008~\citep[not more than $\sim$6~$R_{\rm
s}$;][]{acciari2009}.

\section{Discussion and Summary}
The location and the physical properties of the VHE $\gamma$-ray emission of M~87
are still under hot debate. The core and HST-1 are currently favored as the
emitting sites based on the multi-band correlations detected in the 2005 and 2008
events respectively. In the case of the 2010 flare discussed here,
\textit{Chandra} detected an enhanced X-ray flux from the nucleus during the VHE
flare, while HST-1 remains low state~\citep{harris2011}. This is reminiscent of
the 2008 case. The observed timescales of short variability at VHE/X-ray suggest
the size of the emitting region to be less than $\sim$10$\delta~R_{\rm s}$. These
lead to the argument that the 2010 VHE flare probably originates in the innermost
jet region, at least within the resolution element of \textit{Chandra}
$(0.\hspace{-.3em}^{\prime\prime}6\sim50~\rm{pc}; A12)$.

At the radio bands, we did not find remarkable activities both in the inner jet
and HST-1 during this event. However, the obtained results tend to favor the above
interpretation rather than the HST-1 origin.  For the HST-1 region, we did not
find any compact sub-components that could account for the small emitting volume
suggested from the rapid VHE/X-ray variabilities. From the observed optically thin
radio spectra of HST-1, one can expect simultaneous correlation between radio and
VHE if the 2010 event originates in this feature, but instead the radio flux is
constantly decreasing. The discovery of a new component at the HST-1 upstream in
2010 is particulary intriguing (G12), but the timing of its appearance seems to be
slightly offset from the VHE event. The radio core, on the other hand, remains
very compact during this period. Gaussian model fitting to the 43-GHz visibility
data indicates a deconvolved size of the radio core as $\sim$0.12~mas (17~$R_{\rm
s}$), which is close to the size of $\lesssim10\delta~R_{\rm s}$. Moreover, the
observed optically thick nature of the radio core is possible to hide or weaken
the VHE-related activity at the radio bands, which could be related to the
relatively moderate evolution of the radio core flux. A possibility that the jet
between the core and HST-1 is the $\gamma$-ray emtting source is not completely
ruled out because \textit{Chandra} does not resolve this region. However, this
scenario seems to be problematic because this region is relatively extended
compared with the suggested emitting volume, and maintains the optically thin
radio spectra with the constant flux density during the VHE flare.

It should be noted that the core shift measurements obtained in these observations
locate the 43-GHz radio core position as $\sim$20~$R_{\rm s}$ from the central
engine~(H11). This implies that the site of the VHE $\gamma$-ray associated with
the core is not more distant than $\sim 20~R_{\rm s}$ from the black hole. In this
case, the ambient radiation field such as from the accretion flow could absorb the
emitted TeV photons due to the process of photon(VHE)- photon(IR) pair
creation~\citep{neronov2007}. However, M~87 has only a weak IR nucleus~\citep[$\nu
L_{\nu}\sim10^{40-41}~{\rm erg~s^{-1}}$;][]{perlman2001}, so this allows the
$\gamma$-ray photons up to $\sim$20 TeV to escape even from within 20~$R_{\rm s}$
of the black hole~\citep{brodatzki2011}.

If the VHE flare in 2010 really originates in the core, one should explain the
reason for the distinct behaviour of the radio-to-$\gamma$-ray correlation between
2008 and 2010. One possibility is the difference of the opacity at the radio bands
between 2008 and 2010 due to the change of magnetic field strength. Assuming that
the emitting plasma is spherical and uniformly magnetized, one can estimate the
magnetic field strength through the condition of synchrotron-self-absorption (SSA)
as $B=3.2 \times 10^{-5} \theta^4 \nu_m^{5} S_m^{-2}\delta (1+z)^{-1}$~Gauss,
where $\theta$, $\nu_m$, $S_m$ and $\delta$ give the angular size of the emitting
region in mas, the SSA turnover frequency in GHz, the flux density at $\nu_m$ in
Jy~\citep{kellermann1981}.  Adopting $\theta \sim 0.12$~mas, $\nu_m\gtrsim43$~GHz,
$S_m \sim 0.7$~Jy, $\delta \gtrsim 1$, $B$ is estimated to be $\gtrsim$~2.0~G. In
the case of the 2008 flare, on the other hand, a time-dependent modeling of the
43-GHz core light curve based on SSA indicates a relatively moderate magnetic
field strength as $\sim$0.5~G~\citep{acciari2009}. If we assume that the other
parameters maintain roughly the same values between 2010 and 2008, this yields
$\nu_{m}\lesssim$35~GHz, resulting in the 43-GHz core to be partially optically
thin in 2008. Thus, more amount of the radio emission is possible to escape from
the $\gamma$-ray emitting site, leading to the stronger radio/VHE correlation in
the 2008 case.

The radiative cooling due to synchrotron emission is also possible to contribute
the weaker activity of the radio core in 2010 because of its relatively stronger
magnetic field. The time scale of the synchrotron cooling at a frequency $\nu_b$
can be estimated as $t_{\rm synch} \sim B^{-3/2} \nu_b^{-1/2}$, where $t_{\rm
synch}$, $B$ and $\nu_b$ are measured in year, Gauss and
GHz~\citep{scheuer1968}. The cooling time scale of the 43-GHz emission $t_{\rm
synch, 43GHz}$ under $B\gtrsim2.0$~Gauss results in $t_{\rm
synch}\lesssim$~20~days. This time scale is comparable to the duration of the
possible decaying pattern of the radio core flux (seen for the region (i) between
MJD~55295 and MJD~55332).

While various $\gamma$-ray production models for the black hole vicinity/jet
formation region are proposed~\citep{reimer2004, neronov2007, lenain2008,
tavecchio2008, barkov2010}, it is not easy to discriminate such models in these
observations alone. To constrain the exact location and match the $\gamma$-ray
emission process to a specific model, the use of higher frequency
VLBI~\citep{doeleman2012} is promising, which provides higher transparency to the
$\gamma$-ray emitting region with an event horizon scale resolution.

\acknowledgments

We acknowledge the anonymous referee for his/her careful review and suggestions
for improving the paper. KH thanks R.~C.~Walker for useful discussion in
Tucson. KH is also grateful to S.~Kameno, S.~Mineshige, K.~Tatematsu and
Y.~Sekimoto for useful comments. KH is supported by the Canon Foundation in
Europe. The Very Long Baseline Array is operated by the National Radio Astronomy
Observatory, a facility of the National Science Foundation, operated under
cooperative agreement by Associated Universities, Inc. This work was partially
supported by KAKENHI (24340042 and 24540240). This work made use of the Swinburne
University of Technology software correlator~\citep{deller2011}, developed as part
of the Australian Major National Research Facilities Programme and operated under
licence. Part of this work was done with the contribution of the Italian Ministry
of Foreign Affairs and University and Research for the collaboration project
between Italy and Japan.


\clearpage

\end{document}